\documentclass[aps,prl,
twocolumn,
groupedaddress,showkeys,showpacs]{revtex4-1}

\newcommand{\be}{\begin{equation}}
\newcommand{\ee}{\end{equation}}
\newcommand{\bea}{\begin{eqnarray}}
\newcommand{\eea}{\end{eqnarray}}

\usepackage{graphicx}

\begin{document}


\title{Optical properties and electron transport in low-dimensional nanostructures}


\author{K. Kr\'al}
\email[]{kral@fzu.cz}
\affiliation{Institute of Physics, Academy of Sciences of
Czech Republic, v.v.i. \\ Na Slovance 2, 18221 Prague 8, Czech Republic}

\author{Miroslav Men\v s{\'{\i}}k}
\email[]{mensik@imc.cas.cz}
\affiliation{Institute of Macromolecular Chemistry \\ Academy of Sciences of Czech Republic, v.v.i. \\  Heyrovsk\'y Sq. 1888/2, 16206 Prague 6, Czech Republic}


\date{\today}

\begin{abstract}
We present the theory of the electronic transfer and the optical properties of the quasi-zero dimensional quantum nanostructures, like quantum dots or the DNA molecule. The theory is based on the multiple scattering of the charge carriers in the quasi-zero dimensional nanostructures leading to the manifestation of the nonadiabatic influence of the atomic lattice on the charge carriers. The theory is based on the nonequilibrium Green's functions and the quantum kinetic equations. Three examples of the electronic motion in the small systems are presented, together with a comparison of the theoretical results with their experimental counterparts. The comparison with the experiments underlines importance of the electron-phonon interaction in nanostructures.
\end{abstract}

\pacs{72., 78., 72.80.Ey, 72.80.Le, 73.21.La, 73.63.Kv, 78.67.Hc}
\keywords{quantum dots, electron-phonon interaction, optical properties, electron relaxation, DNA molecule}

\maketitle


\section{Introduction}

The technical problems like those with the development of the heat and in the effort to increase the speed of the information processing in the semiconductor microelectronic circuits have lead to the current trends to achieve the miniaturization of the electronic devices. With respect to the bulk solids, in the systems with a small size the motion of the charge carriers becomes restricted in at least one of the three dimensions. The quantum dots and nanoparticles differ significantly from the bulk and from the higher dimensional nanostructures because the motion of the charge carriers is restricted in all three dimensions. The three-dimensional restriction influences the orbital motion of the charge carriers, the electrons and the holes. For the sake of simplicity we will be speaking only about the electrons in quantum dots.

In quantum dots the spectrum of the stationary bound states of the electrons has a discrete character having usually narrow peaks in the electronic spectral density. This can be observed in optical emission or absorption spectra. The features appear to be seen in the quantum dots made of the materials in which the mean free path of the electron is larger than the lateral size of the quantum dot. The discrete character of the quantum dot electronic energy spectra is therefore found in the inorganic semiconductor materials \cite{Ledentsov1998} and metals \cite{Koh2008}.

The experimental optical spectra sometimes \cite{Fasching} seem to have the form of the spectral lines different from the simple form of the delta-function of the energy variable. The reason for this shape can be connected with the motion of the atomic lattice \cite{jaPRB1998}. Let us briefly show ad hoc arguments supporting this viewpoint: The electrons in a quantum dot necessarily collide with the atomic lattice of the nanostructure. While in the bulk systems the electron leaves the target after the collision and may finish the scattering act upon leaving to infinity, in the nanostructure the process may be different. In the extreme case of the nanostructure confined in all three dimensions, in the quasi zerodimensional (0-D) nanostructure, the electron leaving the target meets with the boundary of the nanostructure a reflects back again to continue the multiple scattering act with the atomic lattice. In this sense the multiple scattering processes should not be a rare phenomenon in the quantum dots.

The multiple scattering process should produce the multiple phonon states. This means that the number of the phonons in a given vibrational mode is then generally different from 0 and 1. The multiphonon states may remind us the coherent photon states of the laser light. The coherent light is in some properties similar to the classical wave of the electromagnetic field, in which, say, the electric field executes classical oscillations in time. In the case when the atomic lattice of the 0-D nanostructure is brought to such a state we may have the system in which the effective Hamiltonian for an electron may, to some extent, depend explicitly on time and the energy is not then completely conserving. Although the discussion we present is only a qualitative one, we expect that the electrons in the atomic lattice of the 0-D nanostructure do not move adiabatically and the lattice influences their motion in a certain way. The nonadiabaticity is manifested in such a way that the electron and the lattice do not exchange only heat. The execute a force on each other. Because of these circumstances the electron and the vibrations of the atomic lattice need not have the same effective temperatures, in the case when we ascribe a temperature to the separate subsystem of the electrons and to that of the vibrations. From the reason of this serious lack of the complete adiabaticity between the motion of the electrons and phonons, we may quite generally expect that developing the theory of the electronic properties of the nanostructures with vibrations on the basis of an assumption of an overall thermodynamic equilibrium may be sometimes not sufficiently satisfactory.

We shall concentrate our attention on the manifestations of the interaction of electrons with the optical phonons of quasi-zero dimensional nanostructures like those of quantum dots made from the polar semiconductors \cite{japss1997,japss1998,tsuchiya,jaPRB1998,jaarXiv1,stationary2005, jaIEEE2004, jaSurfSci2004} or other quasi-zero dimensional structures like the bases of the DNA molecule. The electron-phonon interaction in quantum dots is an intrinsic mechanism which is always present in these objects. The question about what is the strongest electron scattering interaction influencing the electron kinetics and the optical properties of quantum dots does not seem to be completely clarified until these days. From one side, the number of papers containing interpretations of the experiment on the basis of the electron-electron interaction is extensive \cite{Klimov2003}, but on the other side, there is still a considerable number of the experimentally detected effects which, besides the electron-electron interaction, can be strongly influenced by the electron-phonon interaction as well.

Using three examples, this work aims to bring additional arguments in favor of the importance of the electron-optical-phonon interaction in the quantum dots. The first example will deal with the dependence of the optical line width of the quantum dot optical emission on the lateral size of the dot \cite{Takemoto2000,Masumoto2001}. The second example deals with the long-time luminescence decay in quantum dot samples with the intra-dot electron-phonon multiple scattering \cite{Shamirzaev2003,Shamirzaev2004}. In the third example we deal with a mechanism of the irreversible transfer of a charge carrier between two quasi-zero dimensional nanostructures. We present an explanation of the electric conduction measured earlier in the molecules of DNA \cite{Gruner,Torikai}.

\section{Electron-phonon interaction in a two-level model of single quantum dot}

In the three examples we shall use the simple picture of a single electron moving in an individual quantum dot.  As it was done in the earlier papers \cite{japss1997,japss1998,tsuchiya,jaPRB1998,jaarXiv1,stationary2005, jaIEEE2004, jaSurfSci2004} we shall assume that the holes in the valence band states of the dots are much heavier than the electrons in the conduction band states, so that they can be considered approximately as a static charge which contributes only to the overall potential in which the conduction state electron particles move. We shall also assume that because of the larger effective masses the spectrum of the hole band bound states is so dense that comparing to the light electron the holes relax their energy quickly with help of certain interactions like the electron-acoustic phonons interaction, quickly reach their ground state in the region of the hole states and remain occupying this ground state at low temperatures. From this reason we confine ourselves to putting an emphasis only on the kinetics of the conduction band states electrons in their quantum dot bound states.

As it is well known from the times of the research in the so called hot electrons in the polar bulk semiconductors, the interaction of the conduction band electrons with the longitudinal optical (LO) optical phonons has a strong influence on the electronic motion. Let us thus briefly show the basic properties of the two-electronic level model of an electron interacting with the longitudinal optical (LO) phonons in quantum dots. The details of this argumentation were shown previously \cite{japss1997,japss1998,tsuchiya,jaPRB1998,jaarXiv1,stationary2005, jaIEEE2004, jaSurfSci2004} and will not be repeated here. We shall confine ourselves to the simple electronic model Hamiltonian $H_0=\sum_{n=0,1}E_nc^+_nc_n $ with two nondegenerate electronic bound states energy levels only, neglecting spin, with two orbital motion energies $E_0$ and $E_1$ and corresponding states with indexes $n=0,1$ of the electronic orbital motion.

As the electron-lattice interaction we shall use the well known Fr\"ohlich's coupling operator,
\be
H_1=\sum_{{\bf q}, m,n=0,1} A_q \Phi(n,m,{\bf q})(b_{\bf q}-b^+_{-\bf q})c^+_nc_m,
\ee
between the
electron and longitudinal optical phonons. This operator contains the coupling constant $A_q$, which
is $A_q=(-i e/q)[E_{LO} (\kappa^{-1}_{\infty}-
\kappa^{-1}_0)]^{1/2}(2\varepsilon_0V)^{-1/2}$, where
$\kappa_{\infty}$ and $\kappa_0$ are, respectively, high-frequency
and static dielectric constants, $\varepsilon_0$ is permittivity of
free space, $-e$ is the electronic charge, ${q=\mid {\bf q} \mid}$ is the three-dimensional wavevector of the bulk LO phonon mode
and $V$ is volume of the sample. $\Phi$ is the form-factor,
$\Phi(n,m,{\bf q})=\int d^3{\bf r}\psi^*_n({\bf r})e^{i{\bf
qr}}\psi_m({\bf r})$, taking into account the form of the quantum dot in which the electron moves. Concerning the modes of the optical vibrations in the quantum dot, we assume that the structure of the optical lattice vibrations is practically not touched by the presence of the quantum dot in the whole three-dimensional sample. The Hamiltonian operator of the free LO phonons then is $H_{LO}=\sum_{\bf q}E_{LO}b^+_{\bf q}b_{\bf q}$, in which $E_{LO}$ is the energy of the vibrational quantum of the LO phonons and $b_{\bf q}$ is the corresponding annihilation operator of the Fock's phonon. We shall assume that the bulk LO phonons are dispersionless. The complete Hamiltonian of the two-level single quantum dot then is
\be
H_{1QD}=H_0+H_1+H_{LO}. \label{onedot}
\ee
In the present work we shall use an approximative solution of this Hamiltonian. We shall not make any assumption about an overall thermodynamic equilibrium in the electron-phonon system of the dot.

\subsection{Single quantum dot electron kinetics}

As it has been said in the Introduction, in the quantum dots the energies of the electronic bound states are discrete and the states of the optical vibrations in the Fock's states, with an integer number of the vibrational quanta in a single mode, have discrete energies too. If we wish to consider theoretically the irreversible processes of the electron energy relaxation, we should step beyond the finite order of the perturbation calculation and consider generally the multiple scattering processes leading to vibrational states with a non-integer number of phonons in a single mode \cite{terzi}. The multiphonon states can be viewed as coherent states of the phonon modes. These states are known to have a continuum of energy\cite{terzi}. As such they may allow for the irreversible dissipation processes of the electron energy relaxation within the bound states of a dot connected to an environment, as it has been expected to occur in the electron-phonon theoretical interpretation of the absence of the electron energy relaxation bottleneck in quantum dots\cite{japss1997,japss1998,tsuchiya,jaPRB1998,jaarXiv1,stationary2005, jaIEEE2004, jaSurfSci2004}. Let us remind that in the latter references the optical phonons also play a role of a reservoir keeping at a temperature $T_{LO}$.

Without assuming an overall thermodynamic equilibrium in the quantum dot system we are led to using kinetic equations for the electron-phonon system with the Hamiltonian (\ref{onedot}). With the help of the nonequilibrium Green's functions \cite{LL10} or with using a similar suitable technique \cite{Zubarev71} we can obtain certain characteristics of the optical spectra and the electron kinetics in quantum dots.

We consider the two states of the orbital motion of the electron in a quantum dot, namely the state $n=0$, the ground state of the electron and the state $n=1$, the excited state. We are interested in the time $t$ evolution $dN_1/dt$ of the electron occupation $N_1$ of the excited state $n=1$. Using the so called diagonal approximation, we come to the following formula for the rate of change \cite{jaarXiv1,japss1998} $dN_1/dt$:
\begin{widetext}
\begin{eqnarray}
\frac{dN_1}{dt}
=-\frac{2\pi}{\hbar} \alpha_{01} \left[N_1(1-N_0)\left( (1+\nu_{LO})
\int^{\infty}_{-\infty}dE\,\sigma_1(E)\sigma_0(E-E_{LO})\right.
\right.
\label{rate}
\\
\left.
+\nu_{LO}\int^{\infty}_{-\infty}dE\sigma_1(E)\sigma_0(E+E_{LO})
\right)
\nonumber
\\
- N_0(1-N_1)\left((1+\nu_{LO})
\int^{\infty}_{-\infty}dE\sigma_0(E)\sigma_1(E-E_{LO})
\right. \nonumber
\\  \left. \left.
+\nu_{LO})\int^{\infty}_{-\infty}dE \sigma_0(E)
\sigma_1(E+E_{LO})\right)\right]
. \nonumber
\end{eqnarray}
\end{widetext}
Here $\sigma_0$ and $\sigma_1$ are electronic
spectral densities, $\nu_{LO}$ is Bose-Einstein distribution of LO-phonons at temperature $T_{LO}$ of the lattice and $N_0$ is the average electronic occupation of the electronic state with the index $n=0$. The coupling constant $\alpha_{mn }$ is
\be
\alpha_{mn}=\sum_{\bf q}\mid A_q\mid^2 \mid \Phi (n,m,{\bf q})\mid^2,
~~~~~~\alpha_{mn}=\alpha_{nm}.
\label{alpha}
\ee
It includes the bulk characteristics of the electron-phonon coupling and the form factors $\Phi$ of the quantum dot.

The electronic spectral densities $\sigma_n(E)$ are given\cite{LL10}
by  the retarded self-energy $M_n(E)$ which is calculated from the equation for the electronic self-energy (\ref{Dyson}). For the electronic state $n$, and in the self-consistent Born approximation \cite{japss1997,japss1998,tsuchiya,jaPRB1998,jaarXiv1,stationary2005, jaIEEE2004, jaSurfSci2004}, we have:
\bea
\lefteqn{M_n(E)=\sum^1_{m=0}\alpha_{nm}}
\label{Dyson}
\\ \times
&\{&\frac{1-N_m+\nu_{LO}}{E-E_m-E_{LO}-M_m(E-E_{LO})+i0_+}
\nonumber \\
&+&\frac{N_m+\nu_{LO}}{E-E_m+E_{LO}-M_m(E+E_{LO})+i0_+}\},  \nonumber
\eea
where $N_m$ is again the electronic population of the $m$-th state.

The above outlined equations for the single electron kinetics in the electron-phonon system under consideration allow us to give the theoretical explanation of a number of phenomena observed in quantum dot experiments. It appears that the kinetic processes of the electrons in quantum dots on one side, and the "static" properties of the electrons, as they are given by the optical line shape on the other side, are mutually closely related and cannot be well separated one from another. In particular, the optical line shape is intimately related to the nonadiabatic effect of the upconversion determining a stationary value of the electronic density distributed among the energy levels in a quantum dot.

Besides the effect of the fast electron energy relaxation from the state $n=1$ to the state $n=0$, the equations allow also for the effect of the electronic up-conversion, namely, for a spontaneous promotion of the electronic density to the upper excited state. This means that preparing at the beginning the dot in the state with one electron in at the level $n=0$, one can find at later times a nonzero occupation of the excited state $n=1$. This effect was presented earlier at the references e.g. \cite{jaIEEE2004,jaSurfSci2004}. The kinetic equation (\ref{rate}) can be solved to obtain the time development of the electronic occupation of the two electronic levels at a given temperature of the lattice of the LO phonons, $T_{LO}$. The effect of the steady state occupation of the electronic excited state, different by the nature from the thermal up-conversion of the electronic distribution, is possibly well manifested in the observed lasing of the quantum dot lasers from the higher excited states (see e.g. \cite{amarcus2003}). In this way the nonadiabatic influence of the lattice vibrations on the electrons appears to represent an intrinsic property of the quantum dot nanostructure having a possible impact on the functioning of the quantum dot lasers.

\subsection{Example 1: The optical linewidth in a single dot}
Let us remind once again that we ignore the contribution of the holes in the valence band states of the quantum dot exciton and ascribe the linewidth properties of the exciton emission peaks as solely due to the electrons in the conduction band states of the dot. By doing this we shall avoid the complications connected with considering the whole exciton particle containing the electron and the hole.
\begin{figure}
 \includegraphics[width=7.7cm]{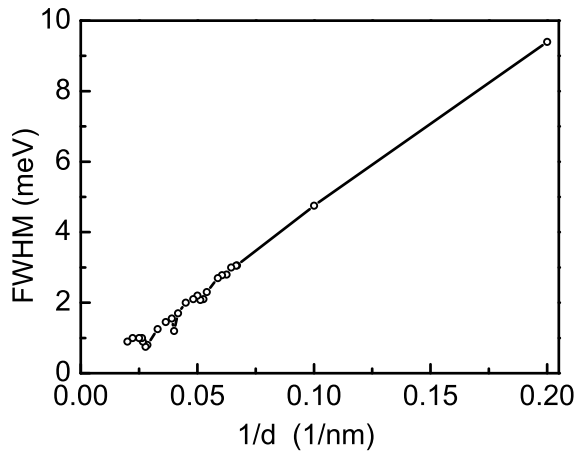}%
\caption{FWHM of the main peak electronic
spectral density in InAs quantum dot, at 10 K plotted
against the inverse of the quantum dot lateral size d.
The result is calculated at the stationary distribution of the
electronic density among the levels. \label{InAs-gamma-1vsd-2}}
\end{figure}
According to the presently used theoretical apprach to the electron-LO-phonon interaction and to the effect of the electron occupation upconversion, it seems that this effect has a direct impact on the properties of the optical transitions linewidth, as it varies with the quantum dot lateral size. From the experimental point of view, about a decade ago measurements were published due to which the full width at half maximum (FWHM) of the lowest energy optical transitions in the quantum dots depends linearly on the inverse value of the quantum dot lateral size $d$. The observed linear dependence goes through the beginning of the coordinates, namely through the point of the width equal zero and the limit of $1/d$ equal zero too. The experiments were performed on the nanoparticles of CdSe, CuCl and CuBr. The measurement technique was the accumulated photon echo experiment \cite{Takemoto2000,Masumoto2001}. The optical linewidth represented by the electronic spectral density line width has been calculated recently to obtain the dependence on the lateral size $d$ (see e.g. Reference \cite{KralMensikGranada}). The calculated linewidth dependence was found for the nanoparticles of InAs semiconductor and also for CdSe \cite{ICTON2009}. In the case of InAs the theory clearly gives the linear dependence found in experiments and goes through the beginning of the coordinates as well (the linewidth is directly proportional to $1/d$). In the course of the theoretical evaluation it has appeared that a key step to be made in the calculation is to fully respect the tendency of the system to achieve the steady state electronic level occupation upconversion. The calculated dependence has come out as a linear function. In the case of the material CdSe the calculation shows that one gets approximately a linear dependence, and the line has the tendency to go through the beginning of coordinates, nevertheless the linear dependence is modified by a certain noise. This computational noise structure is at present attributed to the used approximation. Namely, the noise in the calculated dependence is likely due to the fact that while in the CdSe material the strength of electron-LO-phonon interaction is not small, the calculation uses an approach which is perhaps well suited only for a material with a weak electron-LO-phonon coupling, like for example InAs. The recalculation of the results for the materials with an improved approximation is to be performed.

The Fig.\ \ref{InAs-gamma-1vsd-2} shows a certain fine structure of the width dependence on $1/d$. These features, which are probably difficult to detect experimentally, are numerically obtained at such values of $d$, at which there is a resonance between the electron energy level difference $E_1-E_0$ and an integer multiple of the optical phonon energy.

The direct proportionality of the linewidth to the quantity $1/d$ has not probably been theoretically explained with using another theoretical mechanism so far. The present agreement between experiment and theory provides therefore a supporting argument in favor of an important role of the optical phonons and their multiphonon states in the quasi-zero dimensional nanostructures. It should be emphasized that without taking into account the upconversion effect of the electronic distribution in the dot states, the agreement with experiment in the form of the straight line going through the origin of coordinates would  not be obtained.

\subsection{Example 2: Long-time luminescence decay}
Important applications of quantum dots are connected with a controlled light emission from quantum dots. This application of quantum dots is complicated by the effect of the intermittency of the light emission, or the blinking of the quantum dots \cite{LeeOsborne}, which is so far generally not under control. Under a continuous excitation of the quantum dot sample the emission of a single dot is usually not continuous. According to experimental data the emission is nearly continuous within finite time intervals. The distribution of the occurrence of the lengths of these emission (on state) intervals obeys a power-law statistics \cite{LeeOsborne}. The distribution function of the lengths of intervals, in which a given quantum dot does not emit light (off state intervals) similarly obeys a power-law statistics. Today, a convincing consensus about the mechanism of the blinking seems to be missing.
The quantum dot blinking effect has a certain similarity with another phenomenon observed in the quantum dot samples, namely, with the long-time behavior of the decay of the luminescence intensity signal of a sample after an illumination by a laser pulse \cite{Shamirzaev2003,Shamirzaev2004}. The experiments \cite{Shamirzaev2003,Shamirzaev2004} show that the decay of the luminescence signal does not obey an expected simple exponential law, but it has the form of a power law in a remarkably broad range of the time separation from the exciting laser pulse. In this example of the present paper we show that in a quantum dot sample we can in principle expect a relatively simple intrinsic mechanism which is able to provide the experimentally detected power law decay. As in the previous example the mechanism of this intrinsic effect can be demonstrated upon using the simple two-level quantum dot model.

Let us first remind that the authors of the experiments \cite{Shamirzaev2003,Shamirzaev2004} explain the long-time behavior of the luminescence decay by an assumption that after the laser pulse excitation of the sample a part of the quantum dots capturing the electron and hole carriers are found in the triplet exciton excited state. The quantum dots with triplet a excitation do not emit light and serve a storage of excitation energy for the following period of time. The resulting power-law dependence of the sample luminescence decay is then ascribed to a combination of a large number of channels of charge carrier tunneling from the triplet exciton quantum dots to such quantum dots in which the electron-hole excitation has a singlet exciton character and allows for the emission of light.

In the present example we suggest an intrinsic mechanism which allows to present, in a relatively simple analytical way, a possible origin of the power-law luminescence decay. Let us assume a quantum dot sample consisting of small quantum dots. The size of the dots is such that the singlet-triplet energy separation is larger than the temperature of the experiment. After illuminating the sample by laser pulse, some of the dots with a small diameter contain singlet excitons and some contain the triplet excitons. The triplet exciton dots again serve as a reservoir of the excitations in the sample. The singlet exciton dots relax by emitting light to their ground state typically within about a nanosecond. Let us ignore the processes in the triplet exciton dots which allow the direct emission of light. Let us assume that the triplet exciton dissociates into a hole and electron particles and that the electron is upconverted with the help of the electron-phonon interaction to the excited state in the dot. We also assume that then the upconverted electron immediately leaves the quantum dot and becomes immediately available, without a delay, for a light emission elsewhere in the sample. In other words, after leaving the triplet exciton quantum dot the electron can be captured immediately by another quantum dot in a dense sample of the dots. In the case that the electron meets a hole with a suitable spin in the target quantum dot, then the light is emitted. We shall therefore assume that the key mechanism, or a bottleneck mechanism, determining the time dependence of the long-time luminescence intensity decay is the rate at which the electrons are upconverted from the triplet exciton quantum dot ground state to the excited state in the dot.

Inside a single triplet quantum dot the electron is upconverted to the excited state of the dot by the electron-phonon interaction included in the self-consistent Born approximation to the electronic self-energy, as mentioned above. The key process giving the power-law decay is the rate $dN_1/dt$, given by the formula (\ref{rate}) at which the electron is being promoted to the excited state by the up-conversion. The assumption about the immediate release of the upconverted electron from the triplet exciton dot can be formally represented by the condition that $N_1=0$ for all values of the time variable $t$. Let us note that the effect of the immediate release of the electron from the dot has an important consequence, namely, the total amount of the electron occupation at the triplet exciton dot decreases with time. Because of this decrease, the influence of the total electronic occupation on the multiple phonon generation also decreases, which in turn implies that the upconversion rate becomes slower and in the result obtains a power-law time dependence.

\begin{figure}
 \includegraphics[width=7.7cm]{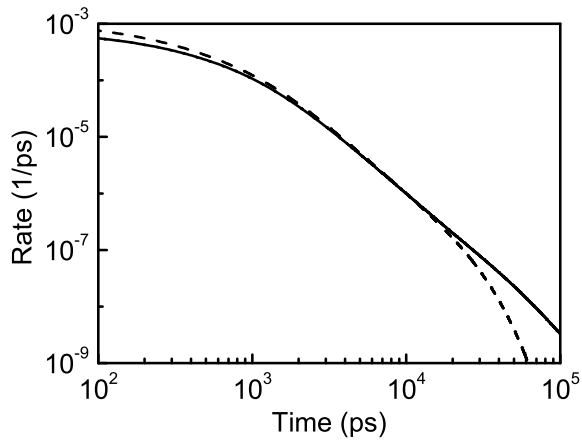}%
\caption{The rate of the upconversion $r=dN_1/dt$ for cubic dot of InAs with the lateral size of 15 nm. $N_1$ is kept equal zero throughout the evolution in time. The full line corresponds to the temperatures 10 K and 30 K, while the dashed line is computed at 50 K. \label{power-line}}
\end{figure}

The power-law decay of the luminescence is shown in the Fig.\ (\ref{power-line}). The model of the electron states used in the present paper in the quantum dot was specified earlier \cite{jaIEEE2004,jaSurfSci2004}. Let us remind at this point that this model uses the method of the effective mass. The electron states are the bound states in an infinitely deep cubic quantum dot. We take into account the ground state of the electron in this potential and one of the triply degenerate lowest energy excited states. In the Fig.\ \ref{power-line} we see that after an initial period of time the relaxation curve signal, being assumed to be proportional to the decaying luminescence signal, becomes a power-law curve. Then for the very long times the curve has a tendency to turn finally to the exponential shape. In our calculation this effect can be due to the damping factors added to the numerical procedure because of the convergence of the calculation.
Fitting the linear part of the obtained curves in the close vicinity of the time of $10^4$ ps to the power law $dN_1/dt=\xi t^{-\alpha}$, we get $\alpha =2.33$.

The purpose of the present luminescence decay example is to show that a simple intrinsic mechanism exists, closely connected to a single quantum dot, which can give a long-time behavior of the quantum dot sample luminescence decay.
Comparing our mechanism with that introduced in refs. \cite{Shamirzaev2003,Shamirzaev2004}, we can say that while the authors of references \cite{Shamirzaev2003,Shamirzaev2004} ascribe the power-law character to a complicated cooperation of various interdot tunneling channels, we conclude that the power-law character of the decay can be obtained already within the electron upconversion inside a single dot.

\section{Two interacting single-level quantum dots}
The charge carrier transfer between neighboring quasi-zero dimensional nanostructures is important for various applications, like the charge transfer along a surface with quantum dots in optoelectronics or even for our understanding the electric conduction of the DNA molecule. In the latter case, we shall discuss the earlier experiments performed by microwave measurements on the DNA molecules, with the result saying that in the limit of low temperatures, below about 100 K, the electric conduction of DNA molecule becomes activationless \cite{Gruner,Torikai}. We show that the property of DNA molecule conduction of being independent of temperature at the low temperatures can be explained by the interaction of the conduction electrons (holes) with the molecular vibrations of the DNA bases in combination with the electron tunneling between the individual quasi-zero dimensional units. The purpose of this demonstration is to show that the interaction of the charge carriers, localized within the molecular bases, with the intra-bases atomic vibrations displays a key role in the determination of the basic properties of this basic biological molecule. In this section we therefore do not pay attention to an electron transfer from one quasi-zero dimensional unit to another with using the presence of the wetting layer, as in the Stranski-Krastanow method grown samples, or the transfer through the electron states in the matrix material of the quantum dot sample.

In order to utilize the theoretical tools developed earlier for the quantum dot systems, we shall develop the theoretical argumentation with using the concepts usual in the quantum dot systems. Using an analogy between two individual quantum dots and two neighboring DNA bases, we shall make a numerical estimate of an electron irreversible transfer process in the DNA molecule. For this purpose the available values of the parameters of the Hamiltonian will be used \cite{Orlandi,Conwell1,Conwell2}.

In a sample of interacting quantum dots the electric current can be transported by the free charge carriers. The electron transport process may be represented by a transfer of electrons within a pair of quantum dots. We choose a simple model consisting of two quantum dots, $A$ and $B$. Each quantum dot is supposed to have only a single electronic orbital for a spinless electron. We shall ignore the electron-electron interactions. Without considering the electron-phonon interaction, the effective Hamiltonian for a conduction band electron can be as follows:
\begin{equation}
H_e=H_{eA}+H_{eB}+V_t.
\end{equation}
Here
\be
H_{eA}=E_Ac^+_Ac_A,
\ee
and
\be
H_{eB}=E_Bc^+_Bc_B,
\ee
are the Hamiltonian operators of an electron on two electronic orbitals, each being localized at a single quantum dot, $A$ and $B$, with the site energies at the dots, $E_A$ and $E_B$. The operator $c_A$ is annihilation operator of electron at the orbital localized at the dot $A$, while the particle operator $c_B$ annihilates electron at the quantum dot $B$. The operator $V_t$ expresses a tunneling of the electron
between the two sites and will have the form
\be
V_t=t(c^+_Ac_B+c^+_Bc_A).
\ee
The above given Hamiltonian assumes that the Hilbert space of the single electron is determined by two electronic orbitals, each being localized at a separate quantum dot. These orbitals will be assumed energetically nondegenerate and will be assumed to have zero mutual overlap.

An electron placed into the system of the two quantum dots is assumed to interact with the vibrations of the quantum dot lattice. Because we assume that each quantum dot has only a single electronic orbital, we consistently choose the electron-phonon interaction in the following form:
\begin{equation}
H_{e-p}=H_{e-p,A}+H_{e-p,B}.
\label{transverse}
\end{equation}
$H_{e-p,A}$ is the operator of the electron-phonon interaction in the quantum dot $A$. In analogy with the earlier theory of the electron energy relaxation in a single quantum dot, and taking into account that there is a single electron orbital per dot, we take this operator in the form taken over, including the notation, from the reference \cite{jaPRB1998}:
\begin{equation}
H_{e-p,A}=\sum_{{\bf q}\in \Omega_A}A_q\Phi_A(0,0,{\bf q})(b_{\bf q}-b^+_{-{\bf q}})c^+_Ac_A.
\end{equation}
Here $A_q$ is Fr{\" o}hlich's coupling constant \cite{jaPRB1998}, given by the material constants of the material of the quantum dots. The integration over ${\bf q}$ covers the range $\Omega_A$ of the phonon wavevector in the quantum dot $A$. The quantity $\Phi_A(0,0,{\bf q})$ is the form-factor of the quantum dot $A$, taking into account the quasi-zero dimensional shape of the quantum dot. Let us remind at this place that we approximate the phonon system of a given dot by the bulk phonon modes of the whole sample \cite{jaIEEE2004}. The sum over ${\bf q}$ represents the sum over the bulk longitudinal optical (LO) modes of the lattice vibrations of the sample. The operator $b_{\bf q}$ is annihilation operator the LO phonon in the mode with the wavevector ${\bf q}$.

Similarly, for the quantum dot $B$ we have the electron-phonon operator as follows:
\begin{equation}
H_{e-p,B}=\sum_{{\bf q'}\in \Omega_B}A_{q'}\Phi_B(0,0,{\bf q'})(b_{\bf q'}-b^+_{-{\bf q'}})c^+_Bc_B.
\end{equation}
Here again $\Phi_B(0,0,{\bf q'})$ expresses the form-factor of the quantum dot $B$ at the value of the phonon wavevector ${\bf q'}$. We consider the phonon modes with the wavevectors ${\bf q}$ and ${\bf q'}$ as simply two different and independent phonon mode systems.

To make the total Hamiltonian complete, we have the noninteracting LO phonon modes Hamiltonian as follows:
\begin{equation}
H_p=\sum_{{\bf q}\in \Omega_A}E^{(A)}_{LO}b^+_{\bf q}b_{\bf q}+\sum_{{\bf q'}\in \Omega_B}E^{(B)}_{LO}b^+_{\bf q'}b_{\bf q'}.
\end{equation}
This free phonon Hamiltonian operator consists of two parts, one is for the phonons of the quantum dot $A$, and the other for the phonons of the quantum dot $B$. We shall assume that the optical phonon energies, being dispersionless in both the quantum dots, are simply equal, $E^{(A)}_{LO}=E^{(B)}_{LO}=E_{LO}$.

The whole Hamiltonian operator $H_{2QD}$ of the system of two interacting quantum dots then is:
\be
H_{2QD}=H_e+H_p+H_{e-p}.
\label{celkovy}
\ee
Because of the boson nature of the phonons we can expect that even if the phonon subsystem is noninteracting with the electrons, the whole system can have a continuum in its energy spectrum. This expectation is obviously due to the possibility of the coherent (multiphonon) states of the individual vibrational modes\cite{terzi}. In the present case the multiphonon states will be present as virtual states brought about by using self-consistent Born approximation to the electronic self-energy \cite{jaPRB1998,jaIEEE2004}.

The Hamiltonian (\ref{celkovy}) is actually one of the forms of the spin-boson Hamiltonian\cite{Leggett87}. We shall obtain the quantum kinetic equations in a perturbative way, using the nonequilibrium Green's functions \cite{LL10}, without making an assumption about the overall thermodynamic equilibrium.
\begin{figure}
 \includegraphics[width=7.7cm]{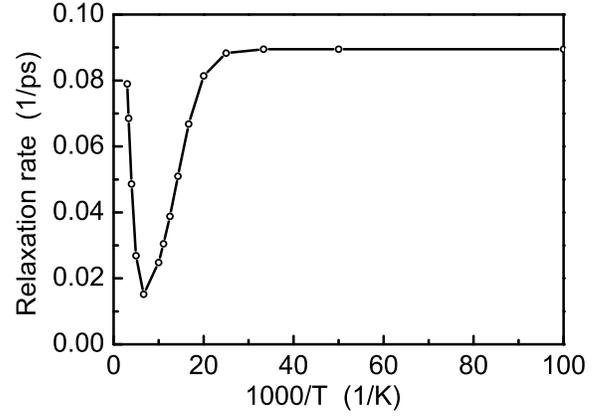}%
\caption{The generation rate of the electronic occupation of the quantum dot $A$, $dN_1/d\tau$ at the state at which the electron occupies the quantum dot $B$ with the lower orbital energy. $T$ is the temperature of the lattice. The quantum dot $A$ has the lateral size of 21 nm, while the quantum dot $B$ has the lateral size of 19 nm. The material parameters of the crystalline GaAs are used for the electron-LO-phonon interaction. \label{rateontemperature-2}}
\end{figure}

\subsection{Example 3: Electron transfer between two quasi-zero dimensional nanostructures}

Although the purpose of this third example is to present a mechanism of the charge transfer between two quasi-zero dimensional elements of a more general type, in choosing the parameters for the Hamiltonian we shall be also motivated by the known electronic structure of the DNA molecule \cite{Orlandi,Conwell1, Conwell2}. We shall assume that the bottleneck part of the DNA molecule, from the point of view of the electron transport, will be such a pair of the neighboring DNA bases, in which an electron is to overcome a potential barrier of the large difference of the electronic site energies on the neighboring bases, $E_A-E_B \approx 0.3\,\, $eV. In an agreement with the well known estimates of the parameters in DNA \cite{Orlandi, Conwell1, Conwell2} we choose that $\mid (E_A-E_B)/t \mid \gg $1, and the inter-bases parameter is taken as $t=0.03\,\,$eV. The molecular bases are represented here approximately by quantum dots which are given by the same model as in the first example, with the cubic shape and the lateral size of 19 nm. The electron wave functions localized within individual molecular bases are chosen to be the ground state wave function in the cubic dot. The electron-phonon interaction in the molecular basis is taken over from the quantum dot system. The details of the parameters of the electron-LO-phonon interaction in quantum dots can be found in references \cite{japss1997,japss1998,tsuchiya,jaPRB1998,jaarXiv1,stationary2005, jaIEEE2004, jaSurfSci2004}. Assuming the electron-phonon interaction as a small perturbation, we perform a canonical transformation of the Hamiltonian diagonalizing the purely electronic part of it. After this operation the Hamiltonian becomes formally identical with the previously used \cite{japss1997,japss1998,tsuchiya,jaPRB1998,jaarXiv1,stationary2005, jaIEEE2004, jaSurfSci2004} two level quantum dot Hamiltonian with the electron-LO-phonon interaction. The electron transfer is thus numerically solved in the same way as it has been done in the case of the single two-level quantum dot, namely including the electron-phonon coupling again in the self-consistent Born approximation to the electronic self-energy. The details of this procedure will be published elsewhere.

We assume that the electric conduction measured in the experimental paper \cite{Gruner} is determined mainly by the bottleneck basis pairs specified above. Calculating the temperature dependence of the rate in time $\tau$, $dN_1/d\tau$, of the electronic transfer from a basis with the lower electron orbital energy $E_B$ to the neighboring basis with the high orbital electronic energy $E_A$, in the cooperation of the tunneling mechanism $t$ and the electron-phonon interaction on both the individual bases, we take the calculated temperature dependence as proportional to the electron conduction of DNA molecule as it is measured by Gruner et al. \cite{Gruner}.

Fig.\ \ref{rateontemperature-2} shows that at the low temperatures the temperature dependence of the electronic conduction of DNA molecule comes out in the present approach as completely activationless, in a rather good agreement with the experimental paper \cite{Gruner}. The activationless character of the conduction is given by the nonadiabatic influence of the atomic lattice on the conduction electron, or in other words, by the upconversion effect of the electron distribution within the system of the two bases of the DNA molecule. We can say the the activationless shape of the temperature dependence of the electric conduction of DNA at the temperature below about 40 K shows that this object behaves as a small system \cite{Spicka2005}. We confine ourselves to the low temperature part of the calculated dependence, not commenting the shape of the temperature dependence of the conduction at the high temperature region at this work.

\section{Conclusions}
The multiple scattering of the electrons of the low-dimensional nanostructures manifests itself significantly in a variety of effects in the electron system in these objects. The comparison with the corresponding experimental data supports the importance of the charge carriers multiple scattering on the optical vibrations of lattice. The theoretical methods used here show that one needs to go beyond the finite degree of the perturbation calculation, or beyond the Golden Rule based theoretical tools. This property of the low-dimensional nanostructures undoubtedly brings about a not so high popularity of the present approach to the theory of the small systems.

The second example discusses an intrinsic affect of a single dot, which should be always present in this structure, and which can give a long-time luminescence in the sample. Although the second example does not seem to resolve the quantum dot blinking problem right away, it gives a mechanism with the power-law character of the luminescence decay.
\begin{acknowledgments}
 The authors acknowledge the support from the projects ME-866, OC10007 of M\v SMT and AVOZ10100520.
\end{acknowledgments}


\bibliography{biba2010}

\end{document}